\documentclass[runningheads]{llncs}

\usepackage{url}
\usepackage{amsmath}
\usepackage{graphicx}

\begin{document}



\title{Novel EEG-based BCIs for Elderly Rehabilitation Enhancement}


\author{Aurora Saibene\inst{1,2}\orcidID{0000-0002-4405-8234} \and
		Francesca Gasparini\inst{1,2}\orcidID{0000-0002-6279-6660} \and
		Jordi Solé-Casals\inst{3}\orcidID{0000-0002-6534-1979}}
\authorrunning{A. Saibene et al.}


\institute{University of Milano-Bicocca, Viale Sarca 336, 20126, Milano, Italy \and
	NeuroMI, Milan Center for Neuroscience, University of Milano-Bicocca, Piazza dell’Ateneo Nuovo 1, 20126, Milano, Italy \\
	\email{a.saibene2@campus.unimib.it, francesca.gasparini@unimib.it}
	\and	
	University of Vic-Central University of Catalonia, C de la Laura 13, 08500, Vic, Barcelona, Spain\\
	\email{jordi.sole@uvic.cat}
}


\maketitle

%

\begin{abstract}
The ageing process may lead to cognitive and physical impairments, which may affect elderly everyday life. In recent years, the use of Brain Computer Interfaces (BCIs) based on Electroencephalography (EEG) has revealed to be particularly effective to promote and enhance rehabilitation procedures, especially by exploiting motor imagery experimental paradigms. Moreover, BCIs seem to increase patients' engagement and have proved to be reliable tools for elderly overall wellness improvement.\\
However, EEG signals usually present a low signal-to-noise ratio and can be recorded for a limited time. Thus, irrelevant information and faulty samples could affect the BCI performance. \\
Introducing a methodology that allows the extraction of informative components from the EEG signal while maintaining its intrinsic characteristics, may provide a solution to both the described issues: noisy data may be avoided by having only relevant components and combining relevant components may represent a good strategy to substitute the data without requiring long or repeated EEG recordings. Moreover, substituting faulty trials may significantly improve the classification performances of a BCI when translating imagined movement to rehabilitation systems. \\
To this end, in this work the EEG signal decomposition by means of multivariate empirical mode decomposition is proposed to obtain its oscillatory modes, called Intrinsic Mode Functions (IMFs). Subsequently, a novel procedure for relevant IMF selection criterion based on the IMF time-frequency representation and entropy is provided. After having verified the reliability of the EEG signal reconstruction with the relevant IMFs only, the relevant IMFs are combined to produce new artificial data and provide new samples to use for BCI training.

\end{abstract}

\section{Introduction}\label{intro}
In the last years, the global growth of the elderly population \cite{vancea2016population} \cite{saibene2020addressing} and the increased life expectancy \cite{saibene2019cognitive} have been determining factors to increase the awareness on the impact that ageing has on elderly people in their everyday life \cite{vancea2016population}. In fact, the ageing process may subjectively affect elderly cognitive abilities and introduce motor control impairments \cite{belkacem2020brain}, which could require the intervention of caretakers and rehabilitation procedures, limiting an elderly person autonomy. Assistive technologies have become particularly attractive to enhance elderly people overall wellbeing, to allow them a certain independence and maintain their social connections \cite{vancea2016population} \cite{saibene2019cognitive} \cite{saibene2020addressing}. 

Among the various technological innovations, the Brain Computer Interfaces (BCIs) have proved to be particularly apt to these tasks and found their application in cognitive and motor rehabilitation systems \cite{liu2014tensor} \cite{gomez2016neurofeedback} \cite{carelli2017brain} \cite{belkacem2020brain} \cite{mane2020bci}. In fact, BCIs allow the decoding of brain dynamics in an on-line configuration \cite{wolpaw2002brain} and can be exploited to control heterogeneous systems (e.g., wheelchairs \cite{herweg2016wheelchair}), and provide an instantaneous feedback to their users \cite{mane2020bci}.\\
The most popular method to allow the recording of the BCI input brain signals is electroencephalography (EEG), which provides multivariate time series collected by placing electrodes on the scalp of a subject. Therefore, the EEG signals have proved to be particularly efficient in accessing brain activities and functions by bringing time, space (electrodes) and frequency information of the neuronal signals \cite{lashgari2020data} in a non-invasive way. \\
Regarding the frequency information, the EEG signal is characterized by different frequency bands (or rhythms) that are representative of specific brain dynamics \cite{vaid2015eeg} \cite{wan2019review}. Table \ref{tab:rhythm} provides a brief overview of the EEG rhythms.\\
Notice that the $\alpha$ and $\beta$ frequency bands can be specifically associated to different cognitive and motor functions and thus their dynamic changes can be widely exploited in rehabilitation systems based on motor imagery (MI) tasks \cite{gomez2016neurofeedback} \cite{szczuko2018comparison}. An MI task consists of the imagination of a real movement, like imaging the opening and closing of a hand, and it has been proved that MI practice improve real movements during a rehabilitation process \cite{liu2014tensor} \cite{mane2020bci}.\\
However, controlling a BCI system with MI is difficult and the ability to perform this task varies from person to person. A good control of a BCI is usually achieved when a user reaches the $70\%$ accuracy in the MI paradigm \cite{kaiser2014cortical}. Reaching this level of accuracy may require a long time, however MI tasks usually enhance brain plasticity and providing a feedback to their users could further improve the cognitive, motor and intellectual functions of elderly users, who are particularly affected by brain dynamic changes due to the ageing process \cite{gomez2016neurofeedback}.

Even though and MI-BCI systems based on EEG signals seem to be a promising tool for rehabilitation purposes, there are many challenges that should be considered and that can affect the overall BCI performance.\\
In fact, the EEG signals, that are at the core of these systems, are extremely heterogeneous and easily affected by noise \cite{roy2019deep}. Moreover, collecting an adequate quantity of reliable data to perform the classification tasks involved in the BCI control system is difficult, due to the lack of sufficient time for recording, the possible poor number of subjects and the difficulty of the experimental tasks \cite{zhang2020data}. These issues may lead to a general classification deterioration and thus affect machine learning model performances \cite{luo2018eeg}.\\
%
\begin{table}
  \centering
  \caption{Overview of the rhythms characterizing the EEG signal.}
    \begin{tabular}{ccc}
    \hline
    \textit{Rhythm} & \textit{Frequency range (Hz)} & \textit{Brain dynamic} \\
    \hline
    $\delta$ & $\leq 4$ & sleep \\
    $\theta$ & $4 - 7$ & drowsiness, sleep, emotional stress \\
    $\alpha$ & $8 - 13$ & relaxed while awake \\
    $\beta$ & $13 - 30$ & alertness, thinking, attention \\
    $\gamma$ & $\geq 31$ & intensive brain activity \\
    \hline
    \end{tabular}%
  \label{tab:rhythm}%
\end{table}%
In the EEG domain, attempts to solve the problem of generating new artificial signals have been provided by the data augmentation literature \cite{luo2018eeg} \cite{zhang2020data} \cite{lashgari2020data}. However, the coherence between these artificial data and the brain dynamics recorded by the EEG should be considered and verified.\\
In \cite{dinares2018new}, an interesting data substitution approach is proposed\footnote{The original code is available at \url{https://github.com/ffbear1993/DR-EMD}.}. The authors exploit the recombination of the oscillatory modes, called Intrinsic Mode Functions (IMFs), obtained through Empirical Mode Decomposition (EMD) \cite{huang1998empirical} of the EEG signals to maintain the EEG time and frequency information even in the artificially produced portions of signals.

Motivated by this work and wanting to provide consistent artificial data to be employed in BCI systems without requiring elderly people to sit over long experimental sessions, we propose a novel processing of EEG data for BCI rehabilitation systems.

Therefore, the present paper extends the work provided by \textit{Dinarès-Ferran et al.} \cite{dinares2018new} and focuses on finding significant IMFs for portions of MI signals presenting tasks of interest, from now on called trials. The IMFs are computed with a multivariate extension of the EMD algorithm \cite{rehman2010multivariate}, wanting to maintain the cross-channel interdependence typical of EEG data \cite{chen2017use}. The IMFs are then selected and recombined in order to obtain an unbiased data substitution.

Therefore, our main contributions are (i) the time-frequency representation of the IMFs, on which (ii) the entropy is computed to define the most informative (relevant) IMFs and (iii) the reconstruction of artificial trials by significant IMF combination.\\
These steps should provide artificial EEG signals presenting coherent brain dynamics and thus exploitable in a BCI paradigm.

The work is then organized as follows. Section \ref{sec:methods} presents the datasets and literature methods employed. Section \ref{sec:our_proposal} describes our proposed approach for relevant IMF selection and artificial data generation. Section \ref{sec:results} discusses the obtained results and Section \ref{sec:conclusions} concludes the work.

\section{Methods} \label{sec:methods}
In this section, the datasets employed to test our proposal and the Multivariate Empirical Mode Decomposition (MEMD) \cite{rehman2010multivariate} used for the IMF decomposition of the EEG signals are described.

\subsection{Datasets}
The following procedures have been tested on 2 datasets: (i) the \textit{EEG Simulated} (ES) dataset \cite{gallego2011application} \cite{gallego2015new}, and (ii) the \textit{EEG Motor Imagery BCI} (MIB) dataset \cite{dinares2018new}.

The choice of dataset ES was driven by the necessity of assessing the efficacy of the proposed relevant IMF selection on controlled data. In fact, the ES dataset presents clean and raw (noise affected with SNR from -20dB to 20dB and adding ocular artifacts) simulated EEG signals on 19 electrodes, i.e., \textit{C$\{3,4,z\}$, F$\{3,4,7,8,z\}$, Fp$\{1,2\}$, O$\{1,2\}$, P$\{3,4,z\}$, T$\{3,4,5,6\}$}, and considering the $\alpha$, $\beta$, and $\gamma$ rhythms. 10 trials of 10 s each have been generated.
For further details, please refer to \cite{gallego2011application} \cite{gallego2015new}.

Instead, dataset MIB was chosen due to the fact that it has been employed by the reference work on artificial trial substitution for a BCI paradigm \cite{dinares2018new}.\\
7 healthy males were asked to perform a MI task consisting of left/right wrist dorsiflexion imagined movements. Each subject participated to 2 experimental runs, during each of which 40 tasks of left and 40 tasks of right wrist MI were randomly performed. A trial consisted of 2s of resting time, an acoustic cue for task preparation and 5s of motor imagination.\\
The recorded electrodes were \textit{C$\{1,2,3,4,5,6,z\}$, Cp$\{1,2,5,6\}$, Fc$\{1,2,5,6,z\}$}. Notice that the dataset has been pre-processed by the authors, who used a bandpass ($0.5 - 30$Hz) and a notch ($50$Hz) filter.
For more details, please refer to \cite{dinares2018new}.

\subsection{Multivariate empirical mode decomposition}
In 1998, \textit{Huang et al.} proposed the Empirical Mode Decomposition (EMD) \cite{huang1998empirical}, a signal processing technique that allows the decomposition of a time series into oscillatory modes (the IMFs), with a completely data-driven approach and avoiding the loss or distortion of the data \cite{zeiler2010empirical}. 
In fact, considering a signal $x(t)$, it (i) finds the signal local maxima and minima, (ii) defines $x(t)$ upper and lower envelopes, (iii) computes their mean envelope and (iv) subtracts it from $x(t)$, obtaining the detail $d(t)$. The computation is repeated assigning $d(t)$ to $x(t)$, until $d(t)$ satisfies the IMFs conditions, i.e., its number of zero-crossings and extrema are equal or differ by a unit and its mean envelope is zero. Thus, $d(t)$ is considered as an IMF and $x(t)$ can be reconstructed by summing the obtained $I$ IMFs and a residuum $\epsilon(t)$: $x(t) = \sum_{i=1}^{I}IMF_i(t) + \epsilon_I(t)$.

Therefore, EMD seems to be suitable to deal with non-stationary and non-linear signals \cite{zhao2001mirror} \cite{gallego2011application}, like the EEG ones. However, it works on multivariate signals with a channel-by-channel approach, thus ignoring the cross-channel interdependence which characterizes the EEG signals \cite{chen2017use}.

To address this issue, \textit{Rehman and Mandic} proposed an EMD variation, i.e., the Multivariate Empirical Mode Decomposition (MEMD) \cite{rehman2010multivariate}. MEMD provides IMFs with the same number of oscillations for each channel by exploiting different projections of the $n$-channel signal into a $n$-dimensional space. Thus, given a multivariate signal $\{\boldsymbol{x}(t)\}_{t = 1}^{T} = \{x_1(t), x_2(t), ..., x_n(t)\}$, MEMD:
\begin{enumerate}
	\item Chooses a suitable direction vector $\boldsymbol{v}^{\theta_k} = \{v_{1}^k, v_{2}^k, ..., v_{n}^k\}$, where $k = 1, 2, ..., K$ and $K$ is the total number of direction vectors and $\boldsymbol{\theta}_k = \{\theta_{1}^k, \theta_{2}^k, ..., \theta_{l}^k\}$ are the angles on a $(n - 1)$ sphere along which are defined the direction vectors.
	\item Computes the $k^{th}$ projection of $\{\boldsymbol{x}(t)\}_{t = 1}^T$ along $\boldsymbol{v}^{\theta_k}$, obtaining $\{\boldsymbol{p}^{\theta_k}(t)\}_{k = 1}^K$ for all $k$.
	\item Finds $\{\boldsymbol{t}_{i}^{\theta_k}\}_{k=1}^K$ time instants, which correspond to the $\{\boldsymbol{p}^{\theta_k}(t)\}_{k = 1}^K$ maxima.
	\item Interpolates $[\boldsymbol{t}_{i}^{\theta_k}, \boldsymbol{x}(\boldsymbol{t}_{i}^{\theta_k})]$ to obtain $\{\boldsymbol{e}^{\theta_k}(t)\}_{k=1}^K$ for all $k$.
	\item Estimates the mean envelope for a set of $K$ direction vectors: $\boldsymbol{m}(t) = \frac{1}{K} \sum_{k = 1}^{K}\boldsymbol{e}^{\theta_k}(t)$.
	\item Extracts the detail $\boldsymbol{d}_j(t) = \boldsymbol{x}(t) - \boldsymbol{m}(t)$, where $j = 1, 2, ..., J$ and $J$ is the maximum number of decomposition scales. If $\boldsymbol{d}_j(t)$ meets the IMF conditions, point 1-3 are applied to $\boldsymbol{x}(t) - \boldsymbol{d}_j(t)$, otherwise to $\boldsymbol{d}_j(t)$.
\end{enumerate}

\begin{figure}
\centering
\includegraphics[width=\textwidth]{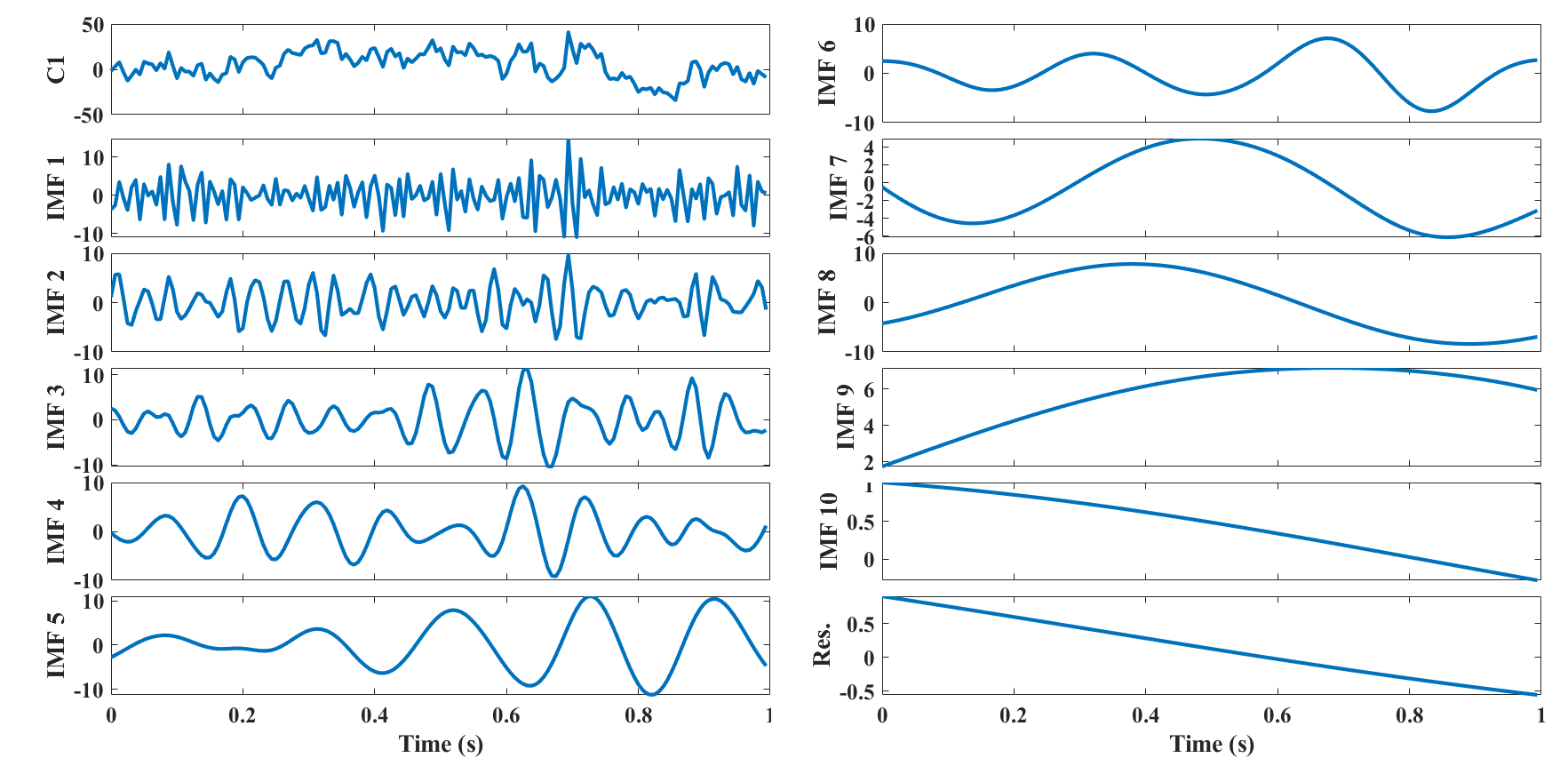}
\caption{Example of IMFs obtained by applying MEMD to electrode \textit{C1}. The last oscillatory mode correspond to the residuum.}
\label{fig:imfimage}
\end{figure}

In this work, MEMD is used to decompose the data of each subject and trial. Fig. \ref{fig:imfimage} shows an example of the IMFs obtained by applying MEMD on the electrode \textit{C1} of a specific subject and trial. The x axis corresponds to the time (s) and the y axis to the signal amplitude ($\mu V$). 
\section{Our proposal}\label{sec:our_proposal}
After having obtained the same number of IMFs for each electrode of a specific subject and trial, the IMFs are used as inputs to the proposed procedure, which consists of 4 main steps: (i) generation of the time-frequency (TF) images for each IMF, (ii) entropy computation on the TF images, (iii) significant IMF selection criterion, and (iv) data substitution.

\begin{figure}
	\centering
	\includegraphics[width=\textwidth]{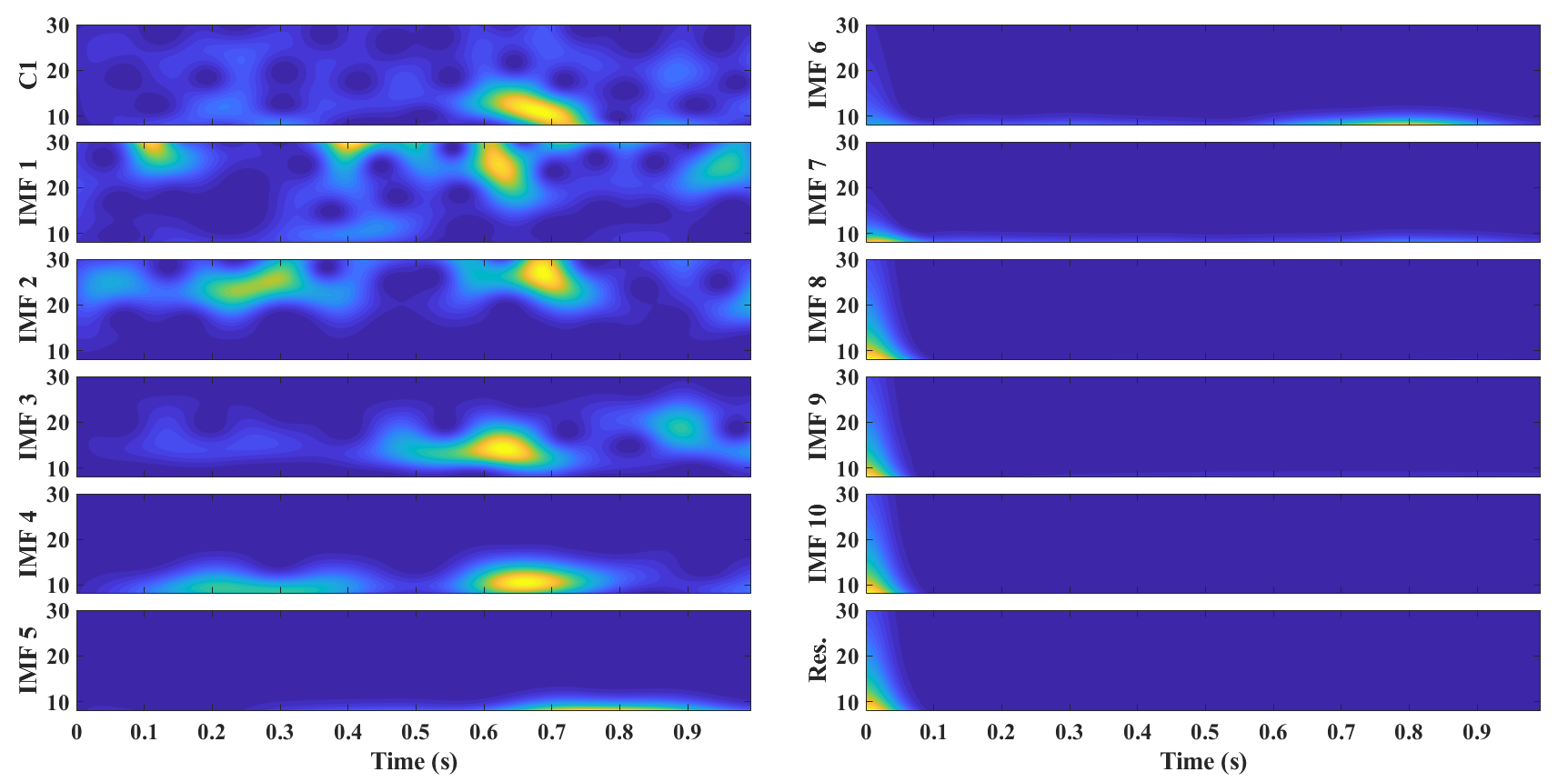}
	\caption{Example of the TF images obtained by applying Morlet wavelet convolution on each of the IMFs presented in Fig. \ref{fig:imfimage}.}
	\label{fig:tfimage}
\end{figure}

Firstly, the TF image generation requires the choice of a frequency range of interest and its time-frequency resolution. Having as targets MI tasks, the chosen frequency range spans from 8 to 30 Hz. In fact, this range includes the $\alpha$ (8-13 Hz) and $\beta$ (13-30 Hz) frequency bands, which are involved in the motor tasks \cite{szczuko2018comparison} (Section \ref{intro}). 
Also, a trade-off between the time and frequency resolution \cite{saibene2020centric} is preferred for image reconstruction.\\ 
Then, the procedure loops on the frequency range and starts with the generation of a complex Morlet wavelet \cite{cohen2019better} exploiting the described parameters. Subsequently, it proceeds with the application of the Fast Fourier Transform (FFT) on both the wavelet and the original signal. Afterwards, the inverse FFT is applied to the wavelet convolved on the signal. Finally, the power data of the convolution is computed and thus the TF image obtained.\\
Fig. \ref{fig:tfimage} presents a colored example of the TF images obtained by computing this procedure on each subject, trial, electrode and relative IMFs (presented in Fig. \ref{fig:imfimage}). The x axis corresponds to the time (s) and the y axis to the frequency range (Hz). 
From the provided example, some clear differences in the IMFs can be observed and we can hypothesize that the IMFs containing more information are the ones which present a more complex texture. Entropy \cite{gonzalez2004digital} can be used to characterize the texture of an image and thus is employed to find the IMFs having more variability. \\
Here the entropy is computed on the TF gray scale image and is equal to $-\sum(p \times \log_2(p))$, where $p$ presents the normalized histogram counts of the TF image.

Considering the entropy obtained on all the IMFs of all the electrodes of a specific trial, the IMFs having entropy greater than the mean entropy are selected as the most significant ones. A more conservative approach is preferred to the elimination of a greater number of IMFs in order to avoid the exclusion of effective neurophysiological signals. The same number of IMFs are selected for all the electrodes of the same trial.

Finally, the last step of our proposal is the artificial data generation, which is performed on each subject by randomly combining the entropy-selected IMFs of different trials of the same task, using a scheme similar to the one described in \cite{dinares2018new} \cite{zhang2019novel}. For each subject, a specific number of artificial trials balanced between the tasks of interest are produced by:
\begin{enumerate}
	\item Finding the maximum number of entropy-selected IMFs $max_{IMF}$ to have the same number of IMFs for each trial. 
	In case a trial presents a lesser number of entropy-selected IMFs if it was decomposed in at least $max_{IMF}$-IMFs, the corresponding oscillation modes discarded by the entropy selection are reintegrated until the trial reaches $max_{IMF}$-IMFs, otherwise IMFs equal to a null vector are added until the trial reaches $max_{IMF}$-IMFs.
	\item Randomly selecting $max_{IMF}$-IMFs from the IMFs of $max_{IMF}$ different original trials for each artificial trial. E.g., a new artificial trial may be composed by the first IMF of the $17^{th}$ original trial (of the same task), by the second IMF of the $5^{th}$ original trial, by the third IMF of the $1^{st}$ original trial and so on until $max_{IMF}$-IMFs are reached.
	\item Reconstructing the artificial trials according to the IMFs obtained at point 2.
\end{enumerate}
Notice that the original trials were reconstructed by considering both their $max_{IMF}$-IMFs to provide coherent signals to compare with the artificial trials and the entropy-selected IMFs. Also, analyses on the trials similarity are performed to ensure the absence of biased artificial trials that could be efficiently used in BCI rehabilitation systems.

\section{Results and Discussion}\label{sec:results}
To better understand the reliability of the proposed relevant IMF selection, a preliminary experiment has been conducted on the ES dataset.

MEMD has been applied for each trial of the clean and raw data, considering all the SNR realizations (from -20dB to 20dB).\\
Subsequently, the TF images of the obtained IMFs have been generated considering the $8 - 30$ Hz frequency range. Finally, the entropy selection criterion has been applied and the entropy-selected IMFs summed up to obtain the reconstructed EEG signal.

The assessment of the proposed method reliability has been performed by computing the similarity between the clean and raw/entropy-reconstructed data by means of Pearson Correlation Coefficient (PCC). The similarity check have been applied to each SNR realization, trial, and electrode.\\
We find that for low SNRs (about -20dB to -17dB), the PCC obtained by comparing the clean data versus the entropy-reconstructed ones do not deviate from the results obtained by applying the PCC on the clean data versus the raw ones. However, a significant increase of the PCC is present for the clean versus entropy-reconstructed signal case with SNR greater or equal to -12dB.\\
Even though the similarity between the clean and entropy-reconstructed signals increases with higher values of SNR, it seems that for the electrodes that are usually affected by ocular artifacts the PCC remains generally low. For all the clean versus raw data cases, the mean similarity remains always lower than 0.9 and for the majority of the electrodes remains under 0.6.

Therefore, the proposed strategy is considered sufficiently reliable for relevant IMF selection and signal reconstruction, having that its results do not deviate from the raw data or that are sufficiently similar to the clean ones.\\


Having obtained a reliable method for relevant IMF selection, it is hypothesized that the entropy-selected IMFs could be efficiently used to reconstruct simulated trials while sufficiently maintaining the intrinsic EEG brain dynamics.

As previously introduced, having that the proposed strategy is modeled on the one described by \textit{Dinarès-Ferran et al.} \cite{dinares2018new} and that considers a BCI experiment, the MIB dataset has been used for testing.\\
The procedure has been computed on each subject and again the MEMD has been applied for each trial separately. Notice that the final goal is to discriminate the left (LW) from the right (RW) wrist MI, which could be applied to control a rehabilitation system.
 
Firstly, random trials were substituted by artificial ones. These trials were obtained by unique combinations of $max_{IMF}$ relevant IMFs selected through the entropy criterion and belonging to $max_{IMF}$ different original trials. 
Remind that $max_{IMF}$ corresponds to the overall \textit{maximum number of IMFs} selected by the entropy criterion. The artificial trials were then reconstructed by summing the selected IMFs.\\ 
To reproduce \textit{Dinarès-Ferran et al.} \cite{dinares2018new} testing, 2.50, 5.00, 7.50, 10.00, 12.50, 25.0, 37.50, 50.00\% of trial substitutions were performed.\\
The Power Spectral Density (PSD) was extracted for each electrode through Morlet wavelet convolution \cite{saibene2021ga}. This feature extraction follows the TF image representation computation, but as a final step, the power data is integrated in the frequency range of interest. Two rhythms were considered separately, i.e., the $\alpha$ and $\beta$ frequency bands, obtaining a total of 38 features. These rhythms have been chosen, due to the presence of MI tasks (Section \ref{intro}). The feature extraction has been restricted on the signal portion during which the MI task is performed to mimic \textit{Dinarès-Ferran et al.} \cite{dinares2018new} experimental setting.\\
Finally, a Linear Discriminant Analysis (LDA) classifier has been applied. For each subject the first run has been used as the training set and the second run as the test set. 

As a first analysis, Table \ref{tab:original_median} reports the median error rates for both the RW and LW conditions after having applied LDA 100 times on the original and reconstructed data.

Trying to mimic the analysis given by \cite{dinares2018new}, the error rate is here intended as the percentage of predicted values that have been wrongly classified for each class.\\
The results obtained by \textit{Dinarès-Ferran et al.} \cite{dinares2018new} (row \textit{DF} of Table \ref{tab:original_median}) have been reported for completeness, however notice that their error rate evaluation is based on all the signal samples and that the features are extracted through common spatial pattern application.\\
The remaining table rows present the results obtained by considering the runs in their original form (row \textit{OO}), and the entropy-reconstructed data of run 1 as the training set and the original data of run 2 as the test set (row \textit{RO}). This last test has been conducted trying to mimic a real-time scenario, during which previously analyzed data (e.g., BCI training phase) may be exploited to predict new unseen data (e.g., BCI translation phase). 

Firstly, notice that the results obtained by the proposed strategy seem to be more balanced compared to the ones reported by \textit{Dinarès-Ferran et al.} \cite{dinares2018new}. Secondly, the \textit{RO} results have been used for comparison with the artificial trial substitution results, which are reported in Table \ref{tab:trial_sub_median}, having that their values are comparable to the ones obtained for the \textit{OO} test.

Notice that the field \textit{AT (\%)} refers to the percentage of trial substitutions balanced between the RW and LW conditions. Therefore, row 1 (\textit{0.00\%}) corresponds to the \textit{RO} results presented in the last row of Table \ref{tab:original_median}.\\ 
Moreover, it can be observed that the results vary from subject to subject. In fact, subject \textit{S01} may be considered a good MI task performer, having that in the RO case the error rates are 2.50 and 0.00 for the RW and LW task, respectively and thus complying with what has been stated in Section \ref{intro}: a person is good in performing MI tasks when he/she can accurately imagine the movement for at least the $70\%$ of the task repetitions. \\
Considering a less stringent constraint, \textit{S07} may be also considered good in the MI experiment. Instead, the remaining subjects seem to have some difficulties performing the MI tasks.\\
Analyzing \textit{S01} and \textit{S07}, besides the 37.50\% and 50.00\% substitutions performed on \textit{S01}'s trial, it can be noticed that the error rates remain stable or improve. Concerning the remaining subjects, the error rates are generally not improved. However, these error rates become more balanced between the conditions and for \textit{S03} and \textit{S04} there seem to be a decrease in the error rate values for some substitution percentages.\\

\begin{table}
	\centering
	\caption{Median error rates (\%) obtained by applying 100 times the LDA classifier on the right (RW) and left wrist (LW) motor imagery.}
	\resizebox{\textwidth}{!}{%
	\begin{tabular}{ccccccccccccccc}
		\hline
		& \multicolumn{2}{c}{\textit{S01}} & \multicolumn{2}{c}{\textit{S02}} & \multicolumn{2}{c}{\textit{S03}} & \multicolumn{2}{c}{\textit{S04}} & \multicolumn{2}{c}{\textit{S05}} & \multicolumn{2}{c}{\textit{S06}} & \multicolumn{2}{c}{\textit{S07}} \\
		\hline
		& \textit{RW} & \textit{LW} & \textit{RW} & \textit{LW} & \textit{RW} & \textit{LW} & \textit{RW} & \textit{LW} & \textit{RW} & \textit{LW} & \textit{RW} & \textit{LW} & \textit{RW} & \textit{LW} \\
		\hline
		\textit{DF} & 5.50   & 6.68  & 11.20 & 66.67 & 29.83 & 20.39 & 42.67 & 32.96 & 36.24 & 35.79 & 27.27 & 39.60 & 58.34 & 22.74 \\
		\textit{OO} & 0.00     & 0.00  & 35.00 & 60.00 & 57.50 & 55.00 & 45.00 & 42.50 & 45.00 & 35.00 & 45.00 & 45.00 & 35.00 & 35.00 \\
		\textit{RO} & 0.00     & 2.50  & 35.00 & 57.50 & 52.50 & 57.50 & 32.50 & 40.00 & 45.00 & 37.50 & 45.00 & 45.00 & 35.00 & 37.50 \\
		\hline
	\end{tabular}%
	}
	\label{tab:original_median}%
\end{table}%

\begin{table}
	\centering
	\caption{Median error rates (\%) obtained by applying 100 times the artificial trial generation and the LDA classifier on the right (RW) and left wrist (LW) dorsiflexion for the trial substitution experiment.}
	\resizebox{\textwidth}{!}{%
	\begin{tabular}{ccccccccccccccc}
		\hline
		& \multicolumn{2}{c}{\textit{S01}} & \multicolumn{2}{c}{\textit{S02}} & \multicolumn{2}{c}{\textit{S03}} & \multicolumn{2}{c}{\textit{S04}} & \multicolumn{2}{c}{\textit{S05}} & \multicolumn{2}{c}{\textit{S06}} & \multicolumn{2}{c}{\textit{S07}} \\
		\hline
		\textit{AT (\%)} & \textit{RW} & \textit{LW} & \textit{RW} & \textit{LW} & \textit{RW} & \textit{LW} & \textit{RW} & \textit{LW} & \textit{RW} & \textit{LW} & \textit{RW} & \textit{LW} & \textit{RW} & \textit{LW} \\
		\hline
		\textit{0.00} & 0.00  & 2.50  & 35.00 & 57.50 & 52.50 & 57.50 & 32.50 & 40.00 & 45.00 & 37.50 & 45.00 & 45.00 & 35.00 & 37.50 \\
		\textit{2.50} & 0.00  & 2.50  & 37.50 & 57.50 & 52.50 & 52.50 & 32.50 & 40.00 & 45.00 & 45.00 & 45.00 & 45.00 & 27.50 & 30.00 \\
		\textit{5.00} & 0.00  & 2.50  & 37.50 & 55.00 & 47.50 & 47.50 & 32.50 & 37.50 & 45.00 & 47.50 & 42.50 & 47.50 & 27.50 & 27.50 \\
		\textit{7.50} & 0.00  & 2.50  & 40.00 & 52.50 & 45.00 & 47.50 & 32.50 & 37.50 & 45.00 & 50.00 & 42.50 & 50.00 & 27.50 & 30.00 \\
		\textit{10.00} & 0.00  & 2.50  & 42.50 & 52.50 & 45.00 & 50.00 & 32.50 & 37.50 & 45.00 & 50.00 & 42.50 & 50.00 & 30.00 & 30.00 \\
		\textit{12.50} & 0.00  & 2.50  & 42.50 & 52.50 & 45.00 & 47.50 & 32.50 & 37.50 & 45.00 & 50.00 & 40.00 & 50.00 & 30.00 & 30.00 \\
		\textit{25.00} & 2.50  & 5.00  & 47.50 & 50.00 & 42.50 & 45.00 & 32.50 & 37.50 & 45.00 & 50.00 & 40.00 & 50.00 & 32.50 & 32.50 \\
		\textit{37.50} & 5.00  & 7.50  & 47.50 & 47.50 & 42.50 & 45.00 & 37.50 & 40.00 & 47.50 & 47.50 & 42.50 & 50.00 & 32.50 & 35.00 \\
		\textit{50.00} & 7.50  & 10.00 & 47.50 & 45.00 & 45.00 & 45.00 & 40.00 & 42.50 & 50.00 & 47.50 & 45.00 & 52.50 & 37.50 & 37.50 \\
		\hline
	\end{tabular}%
	}
	\label{tab:trial_sub_median}%
\end{table}%

Therefore, to ensure that the results obtained with the trial substitutions are coherent with the original results, a double Median Absolute Deviation (MAD) \cite{rosenmai2013using} has been applied to detect if the median results obtained on the original trials could be considered outliers in respect to all the 100 results obtained for each trial substitution.

Notice that if more than the 50\% of the classification results are equal, the MAD is 0. Table \ref{tab:sub_outliers} reports the outliers detected by double MAD application, with the following interpretation: (i) if a cell contains \textit{0 MAD}, it means that the MAD is 0, (ii) if a cell contains \textit{original}, it means that the result obtained on the original trial is considered an outlier in respect to the results obtained on the artificial trial substitution, and (iii) if a cell contains a non-zero number, it means that the double MAD detected that specific number of outliers.  

Analyzing Table \ref{tab:sub_outliers}, the original trial result seems to appear as an outlier in a sufficiently limited number of cases. The proposed strategy results unreliable for \textit{S02}'s RW condition, otherwise it results efficient especially when making few substitutions (2.50 to 5.00\%) or a greater number of substitutions (25.00 to 50.00\%). The overall number of outliers seems also to be fairly low. \\

\begin{table}
	\centering
	\caption{Outlier detection performed through double MAD computation on the original and trial substitution results for the trial substitution experiment.}
	\resizebox{\textwidth}{!}{%
	\begin{tabular}{ccccccccccccccc}
		\hline
		& \multicolumn{2}{c}{\textit{S01}} & \multicolumn{2}{c}{\textit{S02}} & \multicolumn{2}{c}{\textit{S03}} & \multicolumn{2}{c}{\textit{S04}} & \multicolumn{2}{c}{\textit{S05}} & \multicolumn{2}{c}{\textit{S06}} & \multicolumn{2}{c}{\textit{S07}} \\
		\hline
		\textit{AT (\%)} & \textit{RW} & \textit{LW} & \textit{RW} & \textit{LW} & \textit{RW} & \textit{LW} & \textit{RW} & \textit{LW} & \textit{RW} & \textit{LW} & \textit{RW} & \textit{LW} & \textit{RW} & \textit{LW}  \\
		\hline
		\textit{2.50} & 0 MAD & 0 MAD & 0 MAD & 0 MAD & 28    & 9     & 15    & 0 MAD & 6     & 17    & 0 MAD & 0 MAD & 9     & 0 MAD \\
		\textit{5.00} & 0 MAD & 0 MAD & 0 MAD - original  & 11    & 17    & 18    & 0 MAD & 1     & 6     & 9     & 9     & 9     & 9     & 9 \\
		\textit{7.50} & 0 MAD & 0 MAD & 20 & 4     & 1     & 7     & 9     & 0 MAD - original & 10    & 3 - original & 28    & 0 MAD & 10 - original & 10 \\
		\textit{10.00} & 0 MAD & 0 MAD & 17 - original & 8     & 8     & 4     & 16    & 2     & 17    & 4 - original & 28    & 13    & 0 MAD - original & 9 \\
		\textit{12.50} & 0 MAD & 0 MAD & 14 - original & 20    & 7     & 4     & 11    & 0 MAD - original & 13    & 7 - original & 0 MAD & 0 MAD & 0 MAD - original & 3 \\
		\textit{25.00} & 0 MAD & 0 MAD & 17 - original & 1     & 8     & 5     & 3     & 3     & 23    & 1     & 21    & 19    & 4     & 19 \\
		\textit{37.50} & 22 & 0 MAD - original & 15 - original & 8     & 32    & 5     & 4     & 12    & 9     & 1     & 11    & 9     & 12    & 1 \\
		\textit{50.00} & 22 - original & 8     & 13 - original & 1     & 4     & 3     & 20    & 10    & 9     & 2     & 2     & 6     & 14    & 11 \\
		\hline
	\end{tabular}%
	}
	\label{tab:sub_outliers}%
\end{table}%

Making a final evaluation of the results achieved by the trial substitution through the proposed artificial trial generation, the observation given by \textit{Dinarès-Ferran et al.} \cite{dinares2018new} is confirmed: for subjects that naturally perform better the MI tasks, the strategy is generally more effective.\\
In fact, the reported error rates generally remain stable or slightly increase for the other subjects. This effect could be due not only to the difficulties a user may face in performing the MI task which could lead to unreliable trials for a control system, but also to the presence of noisy data.

Therefore, a further development of the trial simulation could be represented by the identification of faulty trials in respect to the experiment of interest. In fact, these trials could deteriorate the overall classification performances and by removing them the trial recombination could benefit the data generation procedure.\\
Moreover, a BCI system could benefit from the faulty trial detection not only intended as noisy trials but also as unreliable trials. In fact, understanding immediately if an elderly person has some difficulties in performing the MI task could provide a better BCI training phase by giving preciser feedback to the subject him/herself and thus increase the task success and decrease subjects' possible frustration.

As a final remark, the proposed methodology seems however to be not extremely faulty when dealing with possibly unskilled subjects. Thus, the performed data substitution may represent a good solution to the EEG data dimensionality problem, providing results that do not deviate excessively from the original ones to train a BCI system.

\section{Conclusions}\label{sec:conclusions}
This work has provided a brief overview of BCI systems based on motor imagery and electroencephalography to enhance elderly people rehabilitation procedures.\\ 
A novel strategy to process the EEG signals before inputting them to the BCI system has been proposed to provide more reliable data without requiring long experimental sessions. Therefore, the EEG signal decomposition by means of MEMD and the relevant IMFs selection through a newly defined entropy criterion have been applied to 2 datasets.\\ 
The proposed approach testing revealed that the signal reconstruction by using only the relevant IMFs is reliable and that the recombination of the relevant IMFs can be efficiently used for artificial trial generation. However, we have noticed that the proposed strategy is particularly suitable for trial substitution of signals recorded from good MI performers, in line with \textit{Dinarès-Ferran et al.} observations.\\
Therefore, we hypothesize that detecting faulty trials in terms of noise and unsuccessful MI performing, may benefit the artificial trial generation as well as the BCI training phase during which an elderly user may effectively improve his/her brain plasticity.\\
Future works will focus on these directions and on exploiting the trial generation strategy for data augmentation.\\
In fact, many data augmentation approaches have been proposed in the literature \cite{lashgari2020data}: additive noise, generative adversarial networks, sliding or overlapping windows, different sampling methods, EEG segment recombination and so on. However, more attention should be required to guarantee the maintenance of the naturally EEG recorded brain dynamics, which we attempted to preserve in the present work.

\bibliographystyle{splncs04}
\bibliography{mybib}


\end{document}